\documentclass[preprint,showpacs,showkeys,preprintnumbers,amsmath,amssymb]{revtex4}

\usepackage{graphicx}
\usepackage{dcolumn}
\usepackage{bm}

\begin{document}

\preprint{APS/123-QED}

\title{Diffusion in Flashing Periodic Potentials}

\author{Bernardo Spagnolo$^{\dagger}$ and Alexander Dubkov$^{\circ}$}
 \affiliation{$^{\dagger}$Dipartimento di Fisica e Tecnologie Relative,
 Universit\`a di Palermo\\ and INFM-CNR, Group of Interdisciplinary
Physics\footnote{Electronic address: http://gip.dft.unipa.it},
Viale delle Scienze, I-90128 Palermo, Italy\smallskip \\
$^{\circ}$Radiophysics Department, Nizhni Novgorod State
University, 23 Gagarin ave., 603950 Nizhni Novgorod,
Russia,\\
$^{\dagger}$spagnolo@unipa.it, $^{\circ}$dubkov@rf.unn.ru}
\date{\today}

\begin{abstract}
The one-dimensional overdamped Brownian motion in a symmetric
periodic potential modulated by external time-reversible noise is
analyzed. The calculation of the effective diffusion coefficient
is reduced to the mean first passage time problem. We derive
general equations to calculate the effective diffusion coefficient
of Brownian particles moving in arbitrary supersymmetric potential
modulated: (i) by external white Gaussian noise and (ii) by
Markovian dichotomous noise. For both cases the exact expressions
for the effective diffusion coefficient are derived. We obtain
acceleration of diffusion in comparison with the free diffusion
case for fast fluctuating potentials with arbitrary profile and
for sawtooth potential in case (ii). In this case the parameter
region where this effect can be observed is given. We obtain also
a finite net diffusion in the absence of thermal noise. For
rectangular potential the diffusion slows down in comparison with
the case when particles diffuse freely, for all parameters of
noise and of potential.
\end{abstract}

\pacs{05.40.-a; 02.50.-r; 05.10.Gg}
\keywords{Brownian motion, Fluctuating supersymmetric periodic
potential, Enhancement of diffusion}

\maketitle

\section{Introduction}

Recently a considerable amount of analysis has been devoted to
investigating transport of Brownian particles in spatially
periodic stochastic structures, such as Josephson
junctions~\cite{Pan04}, Brownian motors~\cite{Rei02} and molecular
motors~\cite{Jul97}. Specifically there has been great interest in
studying influences of symmetric forces on transport properties,
and in calculating the effective diffusion coefficient in the
overdamped limit in
particular~\cite{Rei02,Gan96,Rei01,Mal98,Dub03}. Analytical
results were obtained in arbitrary fixed periodic potential,
tilted periodic potentials, symmetric periodic potentials
modulated by white Gaussian noise, and in supersymmetric
potentials~\cite{Gan96,Rei01,Mal98,Dub03,Fes78,Lin01,ReiHan01}.
The acceleration of diffusion in comparison with the free
diffusion was obtained in Refs.~\cite{Gan96,Mal98,Dub03,ReiHan01}.
At thermal equilibrium there is not net transport of Brownian
particles, while away from equilibrium, the occurrence of a
current (\emph{ratchet effect}) is observed generically.
Therefore, the absence rather the presence of net flow of
particles in spite of a broken spatial symmetry is the very
surprising situation away from thermal equilibrium, as stated in
Refs.~\cite{Rei02,Rei01}. Moreover, the problem of sorting of
Brownian particles by enhancement of their effective diffusion
coefficient has been increasingly investigated in the last years
both from experimental \cite{Gor97,Chi99,Alc04} and theoretical
point of view \cite{Gan96,Bie96}. Specifically the enhancement of
diffusion in \emph{symmetric} potentials was investigated in
Refs.~\cite{Gan96,Alc04}.

Motivated by these studies and by the problem of dopant diffusion
acceleration in semiconductors physics~\cite{Zut01}, we try to
understand how nonequilibrium symmetrical correlated forces
influence thermal systems when potentials are symmetric, and if
there are new features which characterize the relaxation process
in symmetric potentials. This is done by using a fluctuating
periodic potential satisfying the supersymmetry criterion
\cite{Rei01}, and with a different approach with respect to
previous theoretical investigations (see review of P. Reimann in
Ref.~\cite{Rei02}). Using the analogy between a continuous
Brownian diffusion at large times and the "jump diffusion" model
\cite{Lin01,ReiHan01}, we reduce the calculation of effective
diffusion coefficient $D_{eff}$ to the first passage time problem.
We consider potentials modulated by external white Gaussian noise
and by Markovian dichotomous noise. For the first case we derive
the exact formula of $D_{eff}$ for arbitrary potential profile.
The general equations obtained for randomly switching potential
are solved for the sawtooth and rectangular periodic potential,
and the exact expression of $D_{eff}$ is derived without any
assumptions on the intensity of driving white Gaussian noise and
switchings mean rate of the potential.

\section{Fast fluctuating periodic potential}

\begin{figure}[htbp]
\vspace{5mm}
\centering{\resizebox{6cm}{!}{\includegraphics{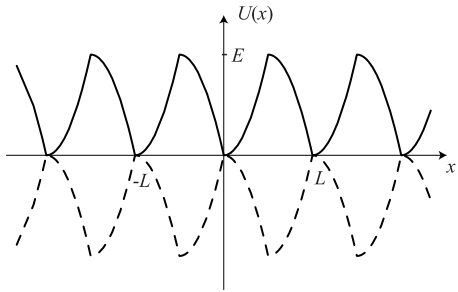}}}
\caption{Periodic potential with supersymmetry.}\label{fig-1}
\end{figure}

\begin{figure}[htbp]
\centering{\resizebox{7cm}{!}{\includegraphics{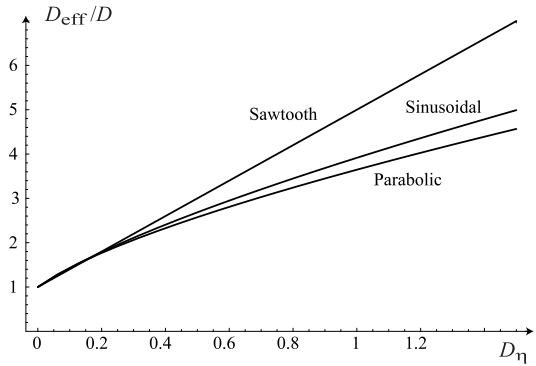}}}
\caption{Enhancement of diffusion in fast fluctuating periodic
potential.}\label{fig-2}
 \vskip-0.4cm
\end{figure}

The effective diffusion coefficient in fast fluctuating sawtooth
potential was first investigated and derived in Ref.~\cite{Mal98}.
In papers \cite{Dub03} we generalized this result to the case of
arbitrary potential profiles. We consider the following Langevin
equation
\begin{eqnarray}
\frac{dx}{dt}=-\frac{dU\left(  x\right)  }{dx}\cdot\eta\left(
t\right) +\xi\left(  t\right) , \label{Lang-2}
\end{eqnarray}
where $x(t)$ is the displacement in time $t$, $\xi\left(  t\right)
$ and $\eta\left(  t\right)  $ are statistically independent
Gaussian white noises with zero means and intensities $2D$ and
$2D_{\eta}$, respectively. Further we assume that the potential
$U\left( x\right) $ satisfies the supersymmetry criterion
\cite{Rei01}
\begin{equation}
E-U\left( x\right) =U\left( x-\frac{L}{2}\right) , \label{SSC}
\end{equation}
where $L$ is the spatial period of the potential (see
Fig.~\ref{fig-1}).
Following Ref.~\cite{Fes78} and because we
 have $\left\langle x\left(  t\right) \right\rangle =0$, we determine the
effective diffusion coefficient as the limit
\begin{eqnarray}
D_{eff}=\lim_{t\rightarrow\infty}\frac{\left\langle
x^{2}(t)\right\rangle }{2t}\, .
 \label{eff}
\end{eqnarray}
To calculate the effective diffusion constant we use the "jump
diffusion" model~\cite{Lin01,ReiHan01}
\begin{eqnarray}
\tilde{x}(t)=\sum_{i=1}^{n(0,t)}q_{i}\,, \label{discrete}
\end{eqnarray}
where $q_{i}$ are random increments of jumps with values $\pm L$
and $n(0,t)$ denotes the total number of jumps in the time
interval $\left( 0,t\right) $. In the asymptotic limit
$t\rightarrow\infty$ the "fine structure" of a diffusion is
unimportant, and the random processes $x\left( t\right) $ and
$\tilde{x}(t)$ become statistically equivalent, therefore
$\left\langle x^{2}\left(  t\right)  \right\rangle \simeq
\left\langle \tilde{x}^{2}(t)\right\rangle $. Because of the
supersymmetry of potential $U\left(x\right)$ the probability
density reads $P\left( q\right) =\left[ \delta\left(  q-L\right)
+\delta\left( q+L\right) \right]/2$. From Eq.~(\ref{discrete}) we
arrive at
\begin{eqnarray}
D_{eff}=\frac{L^{2}}{2\tau}\,, \label{newD}
\end{eqnarray}
where $\tau=\left\langle \tau_{j}\right\rangle $ is the mean
first-passage time (MFPT) for Brownian particle with initial
position $x=0$ and absorbing boundaries at $x=\pm L$. In
fluctuating periodic potentials therefore the calculation of
$D_{eff}$ reduces to the MFPT problem. Solving the equation for
the MFPT of Markovian process $x(t)$ we obtain the exact formula
for $D_{eff}$
\begin{eqnarray}
D_{eff}=D\left[
\frac{1}{L}\int_{0}^{L}\frac{dx}{\sqrt{1+D_{\eta}\left[
U^{\prime}\left(  x\right)  \right]  ^{2}/D}}\right]  ^{-2}.
\label{Main}
\end{eqnarray}
From Eq.~(\ref{Main}), $D_{eff}>D$ for an arbitrary potential
profile $U\left(  x\right)  $, therefore we have always the
enhancement of diffusion in comparison with the case $U\left(
x\right) =0$. We emphasize that the value of diffusion constant
does not depend on the height of potential barriers, as for fixed
potential \cite{Fes78}, but it depends on its gradient
$U^{\prime}\left( x\right)$.
The dependencies of effective diffusion constant $D_{eff}$ on the
intensity $D_{\eta}$ of the modulating white noise are plotted in
Fig.~\ref{fig-2} for sawtooth, sinusoidal and piece-wise parabolic
potential profiles.
\section{Randomly switching periodic potential profile}

Now we consider Eq.~(\ref{Lang-2}) where $\eta(t)$ is a Markovian
dichotomous noise, which takes the values $\pm1$ with switchings
mean rate $\nu$. Thus, we investigate the Brownian diffusion in a
supersymmetric periodic potential flipping between two
configurations $U\left( x\right)  $ and $-U\left( x\right) $. In
the "overturned" configuration the maxima of the potential become
the minima and vice versa. In accordance with Eq.~(\ref{SSC}) we
can rewrite Eq.~(\ref{Lang-2}) as
\begin{equation}
\frac{dx}{dt}=-\frac{\partial}{\partial x}\,U\left(
x+\frac{L}{4}\left[ \eta\left( t\right) -1\right] \right)
+\xi\left( t\right) , \label{shift}
\end{equation}
and the non-Markovian process $x\left( t\right) $ has Markovian
dynamics between flippings. Because of supersymmetric potential
and time-reversible Markovian dichotomous noise the ratchet effect
is absent: $\left\langle \dot{x}\right\rangle =0$. All Brownian
particles are at the origin at $t=0$ and the "jump diffusion"
model (\ref{discrete}) and (\ref{newD}) is used. The probability
density of random increments $q_{i}$ is the same of previous case
and the distribution of waiting times $t_{j}$ reads
\begin{equation}
w\left( t\right) =\frac{w_{+}\left( t\right) +w_{-}\left( t\right)
}{2}\,, \label{wait}
\end{equation}
where $w_{+}\left( t\right) $ and $w_{-}\left( t\right) $ are the
first passage time distributions for the configuration of the
potential with $\eta(0)=+1$ and $\eta (0)=-1$ respectively. In
accordance with Eq.~(\ref{wait}), $\tau $ is the semi-sum of the
MFPTs $\tau_{+}$ and $\tau_{-}$ corresponding to the probability
distributions $w_{+}\left( \tau\right) $ and $w_{-}\left(
\tau\right)$. The exact equations for the MFPTs of Brownian
diffusion in randomly switching potentials, derived from the
backward Fokker-Planck equation, are
\begin{eqnarray}
&&D\tau_{+}^{\prime\prime}-U^{\prime}\left(  x\right)
\tau_{+}^{\prime}
+\nu\left(  \tau_{-}-\tau_{+}\right) =-1\,,\nonumber\\
&&D\tau_{-}^{\prime\prime}+U^{\prime}\left(  x\right)
\tau_{-}^{\prime} +\nu\left( \tau_{+}-\tau_{-}\right) =-1\,,
\label{Hang}
\end{eqnarray}
where $\tau_{+}(x)$ and $\tau_{-}(x)$ are the MFPTs for initial
values $\eta(0)=+1$ and $\eta(0)=-1$  respectively, with the
starting position at the point $x$. We consider the initial
position at $x=0$ and solve Eqs.~(\ref{Hang}) with the absorbing
boundaries conditions $\tau_{\pm}\left(\pm L\right) = 0$. Finally
we obtain the general equations to calculate the effective
diffusion coefficient
\begin{equation}
\theta^{\prime\prime}-f\left( x\right)
\int\nolimits_{0}^{x}f\left( y\right) \theta^{\prime}\left(
y\right) dy-\frac{2\nu}{D}\theta =\frac{xf\left( x\right) }{D}\,,
\label{int-dif}
\end{equation}
\begin{equation}
\frac{D_{eff}}{D}=\left[ 1+\frac{2D}{L}\int\nolimits_{0}^{L}\left(
1-\frac{x}{L}\right) f\left( x\right) \theta^{\prime}\left(
x\right) dx\right] ^{-1}, \label{Accel}
\end{equation}
where $\theta\left( x\right) =\left[ \tau_{+}\left( x\right) -\tau
_{-}\left( x\right) \right] /2$. Equations (\ref{int-dif}) and
(\ref{Accel}) solve formally the problem.

\subsection{Switching sawtooth periodic potential}

\begin{figure}[htbp]
\vspace{5mm}
\centering{\resizebox{7cm}{!}{\includegraphics{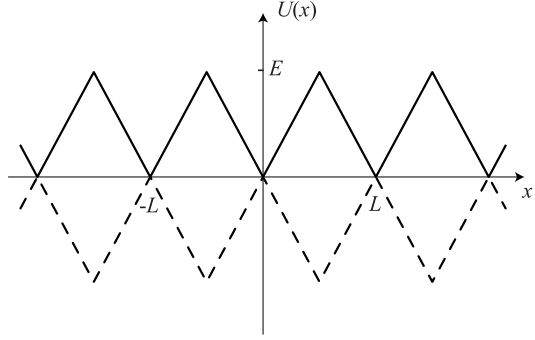}}}
\caption{Switching sawtooth periodic potential.}\label{fig-3}
\end{figure}

\begin{figure}[htbp]
\vspace{5mm}
\centering{\resizebox{7cm}{!}{\includegraphics{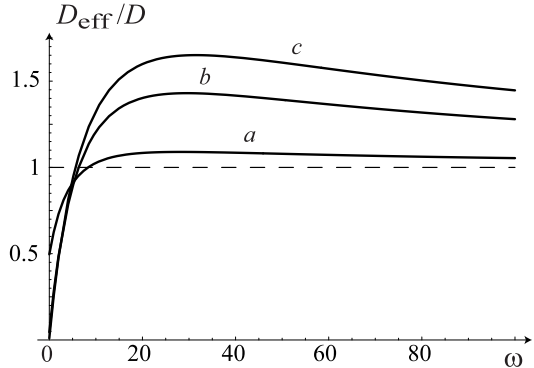}}}
\caption{The normalized effective diffusion coefficient versus the
dimensionless switchings mean rate of potential $\omega =\nu
L^{2}/(2D)$ for different values of the dimensionless height of
the potential barrier. Namely $\beta =3, 7, 9$, for the curves $a,
b,$ and $c$ respectively.}\label{fig-4}
\end{figure}

In such a case (see Fig.~\ref{fig-3}) from Eqs.~(\ref{int-dif})
and (\ref{Accel}), after algebraic rearrangements, we obtain the
following exact result
\begin{equation}
\frac{D_{eff}}{D}=\frac{2\alpha^{2}\left( 1+\mu\right) \left(
A_{\mu} + \mu +\mu^{2}\cosh2\alpha\right)
}{2\alpha^{2}\mu^{2}\left(1+\mu\right) + 2\mu\left(A_{\mu1}\right)
\sinh^{2} \alpha+4\alpha\mu (A_{\mu}) \sinh\alpha+8\left(
A_{\mu2}\right) \sinh^{2}(\alpha/2)} \, , \label{main}
\end{equation}
where $A_{\mu} = 1-3\mu+4\mu\cosh\alpha$, $A_{\mu1} =
7-\mu+2\alpha^{2}\mu^{2}$,  and $A_{\mu2} = 1-6\mu+\mu^{2}$. Here
$\alpha=\sqrt{(E/D)^{2}+\nu L^{2}/(2D)}$ and $\mu=\nu
L^{2}D/(2E^2)$ are dimensionless parameters, $E$ is the potential
barrier height. The Eq. (\ref{main}) was derived without any
assumptions on the intensity of white Gaussian noise, the mean
rate of switchings and the values of the potential profile
parameters. We introduce two new dimensionless parameters with a
clear physical meaning: $\beta=E/D$, and $\omega=\nu L^{2}/(2D)$,
which is the ratio between the free diffusion time through the
distance $L$ and the mean time interval between switchings. The
parameters $\alpha$ and $\mu$ can be expressed in terms of $\beta$
and $\omega $ as $\alpha=\sqrt{\beta^{2}+\omega}$, $\mu=\omega
/\beta^{2}$. Let us analyze the limiting cases. At very rare
flippings $\left( \omega\rightarrow 0\right) $ we have $\alpha
\simeq\beta$, $\mu\rightarrow0$ and Eq.~(\ref{main}) gives
\begin{equation}
\frac{D_{eff}}{D}\simeq\frac{\beta^{2}}{4\sinh^2\left(
\beta/2\right) }\,, \label{rare}
\end{equation}
which coincides with the result obtained for the fixed periodic
potential. For very fast switchings $\left( \omega\rightarrow
\infty\right)$ the Brownian particles ``\emph{see}'' the average
potential, i.e. $\left[ U\left( x\right) +\left( -U\left( x\right)
\right) \right] /2=0$, and we obtain diffusion in the absence of
potential. If we put in Eq.~(\ref{main})
$\alpha\simeq\sqrt{\omega}\left[ 1+\beta^{2}/\left( 2\omega\right)
\right] \rightarrow\infty$ and
$\mu=\omega/\beta^{2}\rightarrow\infty$, we find
\begin{equation}
\frac{D_{eff}}{D}\simeq 1+\frac{\beta^2}{\omega}\,.
\label{large-om}
\end{equation}
The normalized effective diffusion coefficient $D_{eff}/D$ as a
function of the dimensionless mean rate of potential switching
$\omega$, for different values of the dimensionless height of
potential barriers $\beta$, is shown in Fig.~\ref{fig-4}.
A non-monotonic behavior for all values of $\beta$ is observed.
$D_{eff}/D > 1$ for different values above of $\omega$. This value
decreases with increasing height of the potential barrier. In the
limiting case of $ \beta\ll1 $, we find from Eq.~(\ref{main})
\begin{equation}
\frac{D_{eff}}{D}\simeq 1+\frac{\beta^{2}\cdot [\left(
1+2\omega\right) \cosh2\sqrt{\omega} -\left(
4\cosh\sqrt{\omega}-3\right) \left(
1+4\sqrt{\omega}\sinh\sqrt{\omega
}-2\omega\right)]}{2\omega^{2}\cosh2\sqrt{\omega}}\, ,
\label{small-b}
\end{equation}
\begin{figure}[htbp]
\vspace{5mm}
\centering{\resizebox{7cm}{!}{\includegraphics{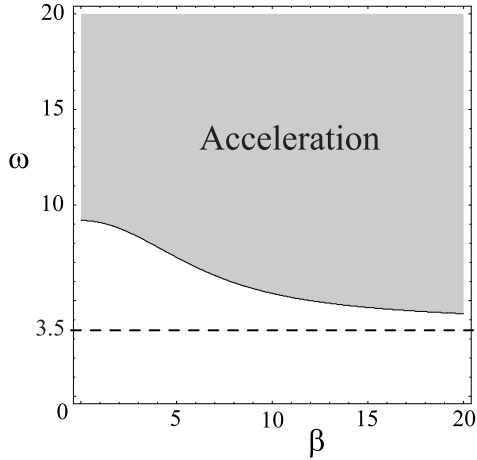}}}
\caption{Shaded area is the parameter region on the plane $(\beta
,\omega )$ where the diffusion acceleration compared with a free
diffusion case can be observed.}\label{fig-5}
\end{figure}
and for low barriers we obtain the enhancement of diffusion at
relatively fast switchings: $\omega >9.195$. For very high
potential barriers $\left( \beta\rightarrow\infty\right)$ and
fixed mean rate of switchings $\nu$, we have
$\alpha\simeq\beta\rightarrow\infty$, $\mu\rightarrow0$, and
$\alpha^{2}\mu\rightarrow\omega$. As a result, we find from
Eq.~(\ref{main})
\begin{equation}
D_{eff}=\frac{\nu L^{2}}{7}\,. \label{mechanics}
\end{equation}
We obtained the interesting result: a diffusion at super-high
potential barriers (or at very deep potential wells) is due to the
switchings of the potential only. According to
Eq.~(\ref{mechanics}) the effective diffusion coefficient depends
on the mean rate of flippings and the spatial period of potential
profile only, and does not depend on $D$. The area of diffusion
acceleration, obtained by Eq.~(\ref{main}), is shown on the plane
$\left( \beta,\omega\right) $ in Fig.~\ref{fig-5} as shaded area.
This area lies inside the rectangle region defined by $\beta>0$
and $\omega>3.5$.

\subsection{Switching rectangular periodic potential}

\begin{figure}[htbp]
\vspace{5mm}
\centering{\resizebox{7cm}{!}{\includegraphics{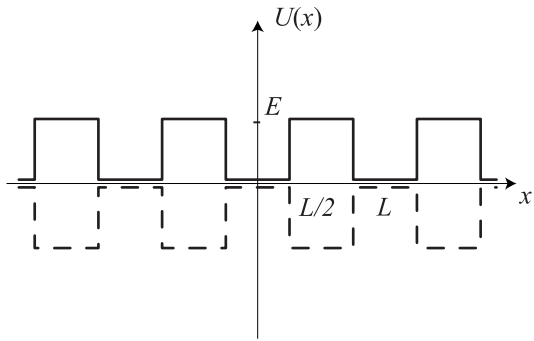}}}
\caption{Switching rectangular periodic potential.} \label{fig-6}
\end{figure}

\begin{figure}[htbp]
\vspace{5mm}
\centering{\resizebox{7cm}{!}{\includegraphics{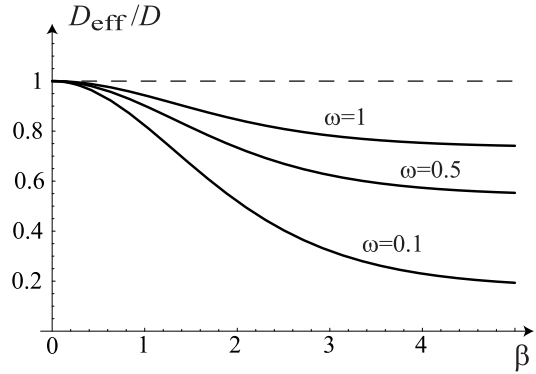}}}
\caption{The normalized effective diffusion coefficient versus the
dimensionless height of potential barriers $\beta =E/D$ for
different values of the dimensionless switchings mean rate
$\omega=\nu L^{2}/(2D)$.}\label{fig-7}
\end{figure}

\begin{figure}[htbp]
\vspace{5mm}
\centering{\resizebox{7cm}{!}{\includegraphics{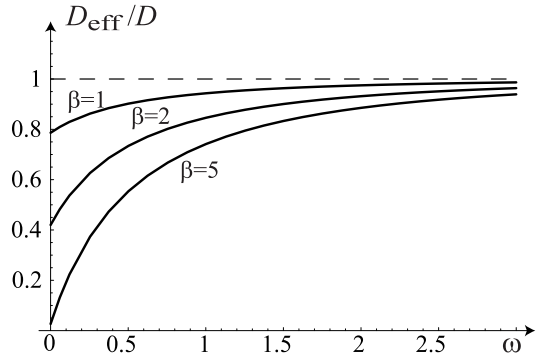}}}
\caption{The normalized effective diffusion coefficient versus the
dimensionless switchings mean rate $\omega=\nu L^{2}/(2D)$ for
different values of the dimensionless height of potential barriers
$\beta =E/D$.}\label{fig-8}
\end{figure}

For switching rectangular periodic potential represented in
Fig.~\ref{fig-6}
the main integro-differential equation (\ref{int-dif}) includes
delta-functions. To solve this unusual equation we use the
approximation of the delta function in the form of a rectangular
function with small width $\epsilon$ and height $1/\epsilon$, and
then make the limit $\epsilon \rightarrow 0$ in the final
expression. As a result, from Eqs.~(\ref{int-dif}) and
(\ref{Accel}) we get a very simple formula
\begin{equation}
\frac{D_{eff}}{D}=1-\frac{\tanh^2{(\beta
/2)}}{\cosh{(2\sqrt\omega)}}\,.\label{rect}
\end{equation}
We have slowing down of diffusion for all values of the parameters
$\beta$ and $\omega$. This is because in rectangular periodic
potential the Brownian particles can only move by thermal force,
crossing randomly the potential barriers as in fixed potential.
The behavior of the normalized effective diffusion coefficient
$D_{eff}/D$ as a function of the dimensionless height of the
potential barrier $\beta$ for different values of the
dimensionless mean rate of switchings $\omega$ is shown in
Fig.~\ref{fig-7}.
The dependence of $D_{eff}/D$ versus $\omega$ for different values
of $\beta$ is shown in Fig.~\ref{fig-8}.
For very rare switchings from Eq.~(\ref{rect}) we obtain the same
result for fixed rectangular periodic potential
\begin{equation}
\frac{D_{eff}}{D}\simeq \frac{1}{\cosh^2{(\beta
/2)}}\,.\label{rarely}
\end{equation}
In the case of very fast flippings the effective diffusion
coefficient, as for sawtooth potential (see Eq.~(\ref{large-om})),
is practically equal to the free diffusion one
\begin{equation}
\frac{D_{eff}}{D}\simeq 1-2e^{-2\sqrt\omega}\tanh^2{(\beta
/2)}\,.\label{rapidly}
\end{equation}
For relatively low potential barriers we get from Eq.~(\ref{rect})
\begin{equation}
\frac{D_{eff}}{D}\simeq
1-\frac{\beta^2}{4\cosh{(2\sqrt\omega)}}\,.\label{low-bar}
\end{equation}
Finally for very high potential barriers $D_{eff}$ depends on the
white noise intensity $D$
\begin{equation}
D_{eff}\simeq \frac{2D}{1+\coth^2{\sqrt\omega}}\,.\label{high-bar}
\end{equation}

\section{Conclusions}

We studied the overdamped Brownian motion in fluctuating
supersymmetric periodic potentials. We reduced the problem to the
mean first passage time problem and derived the general equations
to calculate the effective diffusion coefficient $D_{eff}$. We
obtain the exact formula for $D_{eff}$ in periodic potentials
modulated by white Gaussian noise. For switching sawtooth periodic
potential the exact formula obtained for $D_{eff}$ is valid for
arbitrary intensity of white Gaussian noise, arbitrary parameters
of the external dichotomous noise and of potential. We derived the
area on the parameter plane $(\beta, \omega)$ where the
enhancement of diffusion can be observed. We analyzed in detail
the limiting cases of very high and very low potential barriers,
very rare and very fast switchings. A diffusion process is
obtained in the absence of thermal noise. For switching
rectangular periodic potential the diffusion process slows down
for all values of dimensionless parameters of the potential and
the external noise.

\section*{Acknowledgements}
We acknowledge support by MIUR, INFM-CNR, Russian Foundation for
Basic Research (proj. 05-02-16405), and by Federal Program
"Leading Scientific Schools of Russia" (proj. 1729.2003.2).

\end{document}